\documentclass[showpacs,showkeys,twocolumn,pra,superscriptaddress]{revtex4-1}%
\pdfoutput=1
\usepackage{graphics,amsmath,amsfonts,amscd,revsymb,latexsym,
enumerate,multirow,epsfig}
\usepackage{times}
\usepackage{mathtools}
\usepackage[caption=false]{subfig}
\usepackage{amssymb}
\usepackage{graphicx}%
\providecommand{\U}[1]{\protect\rule{.1in}{.1in}}
\providecommand{\U}[1]{\protect\rule{.1in}{.1in}}

\begin{document}
\title{Leggett-Garg test of superconducting qubit addressing the clumsiness loophole}

\author{Emilie Huffman}
\affiliation{Laboratory for Physical Sciences, College Park, Maryland 20740, USA}
\affiliation{Department of Physics, Duke University, Durham, North Carolina 27708, USA}
\author{Ari Mizel}
\affiliation{Laboratory for Physical Sciences, College Park, Maryland 20740, USA}

\begin{abstract}
The Leggett-Garg inequality holds for any macrorealistic system that is being measured noninvasively.  A violation of the inequality can signal that a system does not conform to our primal intuition about the physical world.  Alternatively, a violation can simply indicate that ``clumsy'' experimental technique led to invasive measurements.  Here, we consider a recent Leggett-Garg test designed to try to rule out the mundane second possibility.  We tailor this Leggett-Garg test to the IBM 5Q Quantum Experience system and find compelling evidence that qubit $Q_2$ of the system cannot be described by noninvasive macrorealism.
\end{abstract}
\maketitle
 
\section{Introduction}
The field of quantum computation has stimulated interest in tests of quantum behavior.  Such tests have provided standardized protocols to showcase control over qubit systems \cite{Ansmann:2009aa,Vlastakis:2015aa}.  They can provide metrics for qubit performance.  Moreover, as a result of experimental advances associated with the quantum computation era, it has become possible to close loopholes in foundational tests of quantum mechanics \cite{Hensen:2015aa,Giustina2015}.

While Bell inequality violations \cite{Bell64} retain their canonical status among tests of quantum behavior, they are ill-suited for many experimental systems.  To apply a Bell inequality test to a system under investigation, the system must possess two parts that can retain quantum coherence while being segregated until they have a spacelike separation.  An alternative to the Bell inequality, one that doesn't make this demand, is the Leggett-Garg inequality  \cite{Leggett1985,Emary2014}.

The Leggett-Garg inequality holds for any macrorealist system that is being measured noninvasively.  It has recently been applied to a number of systems \cite{Palacios10,Xu:2011aa,Athalye2011,Dressel2011,Goggin2011,Waldherr2011,Souza2011,Suzuki2012,Knee:2012aa,George2013,Groen2013,Katiyar2013,Zhou2015,Knee16,Formaggio2016}.  Unfortunately, a clumsiness loophole \cite{Wilde2012} can thoroughly undermine the significance of any violation of the Leggett-Garg inequality.  It is essential to address the clumsiness loophole if one wishes to draw meaningful conclusions from a Leggett-Garg test.

In this paper, we study the IBM 5Q Quantum Experience system \cite{IBM5Q} by carefully implementing a Leggett-Garg program designed to address the clumsiness loophole.  The IBM 5Q Quantum Experience is a publicly accessible system of five superconducting qubits that can be controlled via a website interface.  Earlier papers have exhibited the capabilities of the IBM 5Q \cite{Devitt2016,Alsina2016,Berta2016}.  Our aim is to execute a particularly careful and persuasive demonstration that at least one of the qubits of the IBM 5Q is genuinely quantum, or at least not a macrorealistic system being measured noninvasively.  Our Leggett-Garg test, which is structured to address the clumsiness loophole in the deliberate manner formulated in \cite{Wilde2012}, can also productively inform the design of future tests of other systems.

The paper is organized as follows.  Section \ref{Protocols} frames the six experimental protocols that make up our Leggett-Garg test.  These protocols must be tailored to accommodate constraints in the IBM 5Q system; section \ref{Experiment} describes details.  Results are supplied in section \ref{Results}, and we conclude in section \ref{Conclusion}.

\section{Protocols}
\label{Protocols}
\begin{figure}
\begin{center} 
\includegraphics[width=3in]{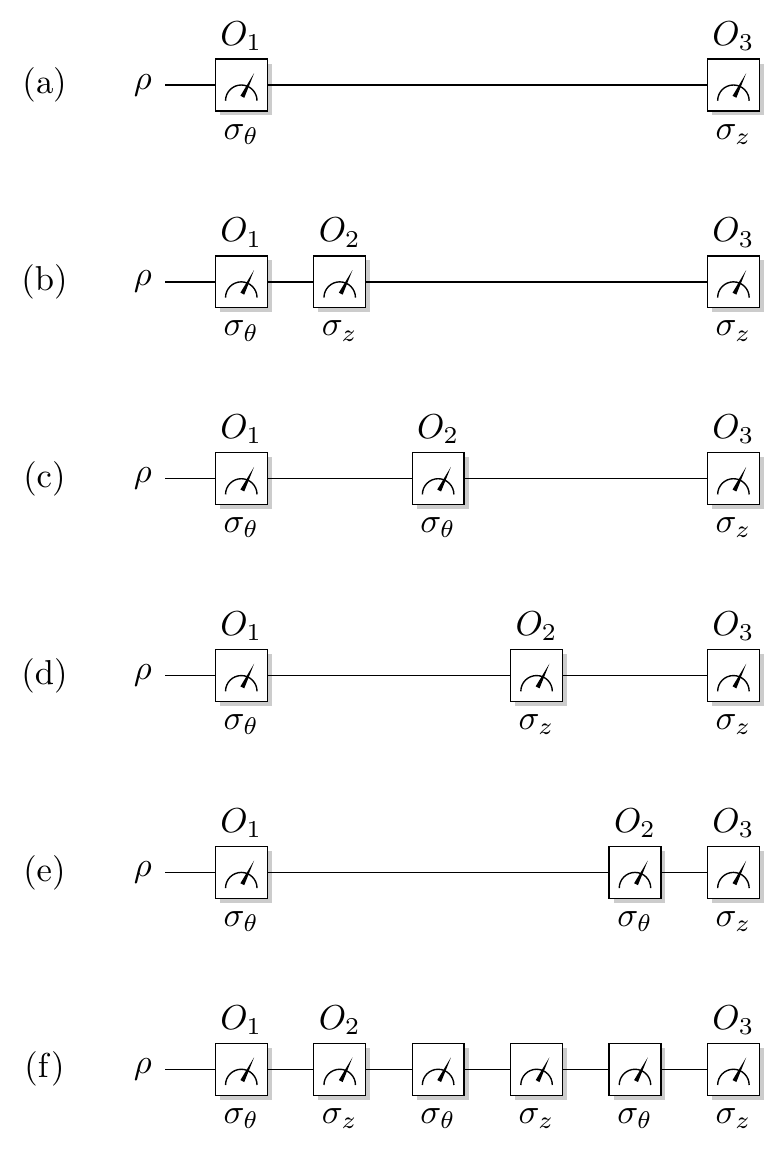}
\end{center}
\caption{Proposed Leggett-Garg experimental program.  The symbols  $\sigma_z$ and  $\sigma_{\theta}$ should be ignored while deriving the Leggett-Garg inequality, which does not presume any quantum mechanics.  The symbol $\sigma_z$ below, say, $O_3$ indicates that, in quantum theory, $O_3$ consists of a measurement along the qubit initialization direction, $\hat{z}$.  The symbol $\sigma_{\theta}$ below, say, $O_1$ indicates that $O_1$ consists of manipulations and a measurement that are equivalent to a measurement along a direction $\sin \theta \hat{y} + \cos \theta \hat{z}$ oriented at an angle $\theta$ with respect to $\hat{z}$.    }
\label{exp}
\end{figure}

We test for violations of macrorealism by adapting the proposal in \cite{Wilde2012}.  Consider an experimental program comprised of the six protocols depicted in Fig. \ref{exp}.  Focus initially on the first and last protocols, (a) and (f), appearing in the figure.

Protocol (a) shows a physical system measured by some operation $O_1$ that is arranged to yield a dichotomous result $1$ or $-1$.  This system is then subjected to operation $O_3$ that yields another dichotomous result $1$ or $-1$.    One can compute the correlator $\left\langle O_1 O_3\right\rangle_a$ by repeating protocol (a) many times and taking the average value of the product  $O_1 O_3$.  

Protocol (f) shows an interleaved series of manipulations and measurements of the two kinds appearing in protocol (a).  In particular, operation $O_2$ is a manipulation and measurement equivalent to $O_3$ but occurring earlier in the series.  A single run of protocol (f) yields measurement results $\pm 1$ for operations $O_1$, $O_2$, and $O_3$.  One can compute the correlators $\left\langle O_1 O_3\right\rangle_f$, $\left\langle O_1 O_2\right\rangle_f$, and $\left\langle O_2 O_3\right\rangle_f $ by repeatedly executing protocol (f), taking the products $O_1 O_3$, $O_1 O_2$, and $O_2 O_3$ each run, and averaging over runs.

For any given run of protocol (f), all 8 possible values of the triplet $(O_1, O_2, O_3) = (\pm 1, \pm 1,\pm 1)$ satisfy the inequality $O_1 O_3 + O_1 O_2 + O_2 O_3 + 1 \geq 0$.  Taking the average of this inequality over repeated runs yields an inequality on correlators $\left\langle O_1 O_3\right\rangle_f + \left\langle O_1 O_2\right\rangle_f + \left\langle O_2 O_3\right\rangle_f + 1 \geq 0$.

Suppose that our physical system is macrorealistic and that all of the operations in protocol (f) measure it noninvasively.  Then,
\begin{equation}
\label{eq:correlatorequality}
\left\langle O_1 O_3\right\rangle_f = \left\langle O_1 O_3\right\rangle_a
\end{equation}
since $O_2$ and the other operations before $O_3$ in protocol (f) do not perturb the system.  Substituting this into our correlator inequality, we obtain the Leggett-Garg inequality
\begin{equation}
LG = \left\langle O_1 O_3\right\rangle_a + \left\langle O_1 O_2\right\rangle_f + \left\langle O_2 O_3\right\rangle_f + 1 \geq 0.
\label{eq:LG}
\end{equation}
If a system violates this inequality, it is not a macrorealistic system undergoing noninvasive measurement.  One exciting possibility is that the system is impossible to describe correctly using any macrorealistic noninvasive theory.  For instance, perhaps the system is quantum mechanical, exhibiting the strange properties described by quantum theory.  But there is a mundane possibility as well.  Perhaps the system is macrorealistic, and can be measured noninvasively, but our measurements are invasive simply because of our experimental clumsiness.  This entirely plausible circumstance is termed the ``clumsiness loophole" in \cite{Wilde2012}.

To address the clumsiness loophole, our full experimental program includes verification protocols (b) - (e) in Fig. \ref{exp} in addition to protocols (a) and (f).  Each protocol (b) - (e) is designed to place a limit, called the $\epsilon$-\textit{adroitness}, on the invasiveness of an operation.
For the $O_2$ measurement in the middle of the experiment in protocol (b), for example, we say that it is $\epsilon_b$-\textit{adroit} if
\begin{equation}
\left|\left\langle O_1 O_3\right\rangle_b - \left\langle O_1 O_3\right\rangle_a\right| \leq \epsilon_b.
\label{eq:epsilonb}
\end{equation}
Similarly, the $O_2$ measurement in protocol (c) is said to be $\epsilon_c$-\textit{adroit} if
\begin{equation}
\left|\left\langle O_1 O_3\right\rangle_c - \left\langle O_1 O_3\right\rangle_a\right| \leq \epsilon_c.
\end{equation}
We define $\epsilon_d$ and $\epsilon_e$ analogously based on protocols (d) and (e).  Assuming that several of these measurements together cannot somehow collude nonlinearly to have an unexpectedly dramatic effect on the system, the maximum effect that the four intermediate measurements in part (f) could have on the correlation function $\left\langle O_1 O_3\right\rangle_f$ is
\begin{equation}
\epsilon_{total} = \epsilon_b + \epsilon_c + \epsilon_d+ \epsilon_e.
\label{eq:epsilontotal}
\end{equation}
By separately testing every single operation that appears in Fig. \ref{exp} protocol (f), we have direct experimental evidence that none of these operations is causing a mundane violation of the Leggett-Garg inequality by clumsy invasiveness.  Designing a Leggett-Garg program with this feature is subtle.

If an experiment yields a value for $LG$ satisfying both
\begin{equation}
LG < 0 \text{ and } \left|LG\right| \geq \epsilon_{total},
\label{eq:bothLG}
\end{equation}
we have evidence that the system can never be correctly characterized by any macrorealistic noninvasive theory.

Suppose that we believe that our system is a qubit correctly described by quantum mechanics.  Will it actually exhibit a violation of eq. (\ref{eq:LG})?  We can derive a quantum mechanical expression for $LG$ in this set of experiments by using the formulae below, where $\sigma_\theta = \sin \theta \sigma_y + \cos \theta \sigma_z$ and the superoperators $\bar{\Delta}$ and $\bar{\Delta}_\theta$, are defined as $\bar{\Delta}\left(\rho\right) = \frac{1}{2}\left(\rho + \sigma_z \rho \sigma_z\right)$ and $\bar{\Delta}_\theta\left(\rho\right) = \frac{1}{2} \left(\rho + \sigma_\theta \rho \sigma_\theta\right)$:
\begin{equation}
\begin{aligned}
\left\langle O_1 O_3\right\rangle_a &= \frac{1}{2} {\rm Tr}\left(\sigma_z, \left\{\sigma_\theta,\rho\right\}\right) \\
\left\langle O_1 O_2\right\rangle_f &= \frac{1}{2} {\rm Tr}\left(\sigma_z, \left\{\sigma_\theta,\rho\right\}\right) \\
\left\langle O_2 O_3\right\rangle_f &= \frac{1}{2} {\rm Tr}\left(\sigma_z \left(\bar{\Delta}_\theta\circ \bar{\Delta} \circ \bar{\Delta}_\theta\right)\left(\left\{\sigma_z, \bar{\Delta}_\theta\left(\rho\right)\right\}\right)\right).
\end{aligned}
\end{equation}
These formulae imply
\begin{equation}
LG = 2 \cos\theta + \cos^4\theta + 1.
\end{equation}
This value is negative if we choose $\theta$ between $.683\pi$ and $\pi$ or between $-.683\pi$ and $-\pi$. We therefore do expect to be able to see a violation of our Leggett-Garg inequality for a qubit.  Note also that protocols (b) - (e) were designed with a qubit in mind such that the intermediate measurements should not change $\left\langle O_1 O_3\right\rangle$, and the $\epsilon$-\textit{adroitness} parameters should be small.  We now tailor this experimental protocol so that it can be implemented on the IBM 5Q.

\section{Experiment}
\label{Experiment}

The IBM 5Q consists of five superconducting transmon qubits patterned on a silicon substrate. The qubits are labeled $Q_0,Q_1,Q_2,Q_3,$ and $Q_4$. There are several constraints on the current qubit setup that are relevant to our proposed experimental program.

\onecolumngrid

\begin{figure}[h]    
    \includegraphics[width=14.8cm]{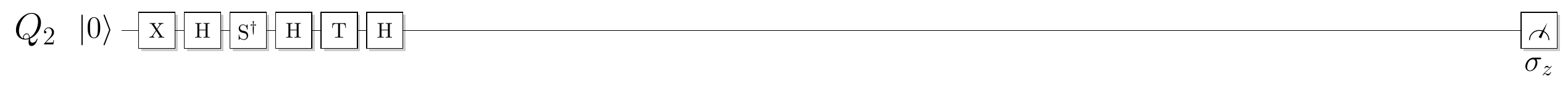}   
    \caption{Protocol (a) of the experiment implemented in a circuit acting on qubit $Q_2$.}
    \label{parta}
\end{figure}

\twocolumngrid

\captionsetup[subfigure]{labelformat=empty}

\onecolumngrid

\begin{figure}[ht]
\centering
		\subfloat[(b)]{
 			\includegraphics[width=14cm]{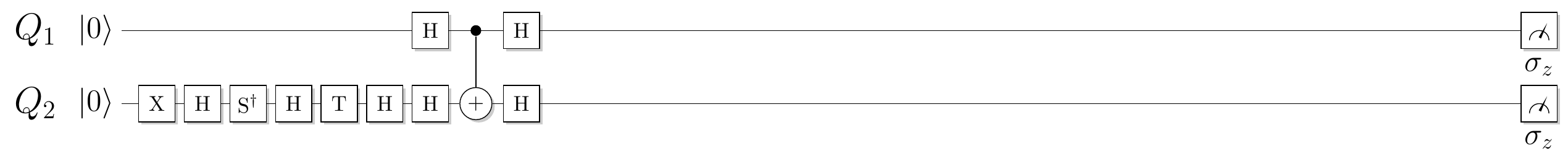} 
		}     
		\vspace{.5cm}
		
		\subfloat[(d)]{
			\includegraphics[width=14cm]{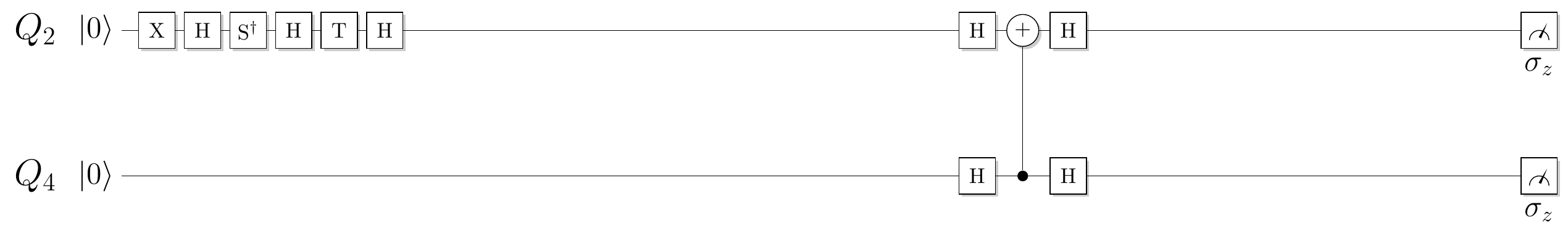}
		}
    \caption{Protocols (b) and (d) of the experimental program implemented in circuits.}
    \label{partbd}
\end{figure}

\twocolumngrid

First, the IBM 5Q permits only one measurement on a given qubit each experimental run.  Fig. \ref{exp} involves multiple measurements on a single qubit.  Rather than performing an additional measurement directly on a single qubit, we therefore perform the measurement by transmitting the qubit's state to an ancilla qubit using $CNOT$ gates and measuring the ancilla qubit.  This is just an alternate realization of the measurement operations in Fig. \ref{exp}; it does not invalidate our carefully constructed Leggett-Garg test.

This modification does force us to consider a second constraint on the IBM 5Q system.  For a 5 qubit system, one might imagine $5 \times 4 = 20$ different types of $CNOT$ gates, targeting any one of the 5 qubits and controlled by any of the remaining 4 qubits.  For the IBM 5Q system, only 4 different types of $CNOT$ gates are available: every $CNOT$ must have $Q_2$ as the target qubit and $Q_0,Q_1,Q_3,$ or $Q_4$ as the control qubit.  To reduce the number of $CNOT$ gates necessary for our experimental program, we choose $Q_2$ to play the role of the qubit that appears in Fig. \ref{exp} and the other qubits as the internal degrees-of-freedom of the measurement devices in Fig. \ref{exp}.

The third and final constraint we consider arises from the fact that there are only five qubits in the IBM 5Q.  Since each qubit can be measured at most once, any IBM 5Q circuit can only make five total measurements. Protocol (f) of Fig. \ref{exp} involves six measurements. To deal with this issue, we treat $O_1$ in Fig. \ref{exp} not as a measurement but instead as an initialization into $\rho_\theta = \left|1 \right\rangle_\theta \left\langle 1\right|_\theta$, a state that should give result $+1$ when subjected to the $O_2$ measurement in protocols (c) or (e) of Fig. \ref{exp}.  (If one wishes to make this as parallel as possible to our discussion of the experimental protocol above, it may be helpful to think of initialization as measurement followed by a postselection on result $1$ rather than $-1$.)  The IBM 5Q device initializes qubits into state $\left| 0 \right\rangle$, so our $O_1$ will consist of a $NOT$ gate followed by a rotation gate.  We note that the correlation function in part (a) for $\rho_\theta = \left|1\right\rangle_\theta \left\langle 1\right|_\theta$ may be rewritten as
\begin{eqnarray}
\left\langle O_1 O_3\right\rangle_a & = & p_{-1,-1} + p_{1,1} - p_{-1,1} - p_{1,-1} \nonumber \\
& = & p_{1,1} - p_{1,-1}  =   \left\langle O_3\right\rangle_a,
\end{eqnarray}
with similar results for every correlator that involves $O_1$.  The significance of the Leggett-Garg conditions (\ref{eq:bothLG}) and the quantum mechanical predictions are unchanged by this alternation in our experimental program.

It turns out to be convenient to choose $\theta=-3\pi/4$ for our $\theta$-measurements in Fig. \ref{exp}.  The IBM 5Q currently permits single qubit gates $X,Z,Y,H,S,S^\dagger,T,T^\dagger$ and measurement in the $z$-direction. To perform a $\theta=-3\pi /4$ measurement, we thus rotate the basis noting that the rotation matrix for $\theta=-3\pi / 4$ obeys the identity
\begin{equation}
e^{-i3\pi \sigma_x / 8} = H e^{-i3\pi \sigma_z/8} H.
\end{equation}
This product has the form $e^{-3i\pi/8}  H T^3 H = e^{-3i\pi/8} H T S H$.  It turns out that the IBM 5Q system exhibits better performance on our experimental program if we re-express the product as $e^{-i\pi/8} H T H S^\dagger H S^\dagger$ using the identity $e^{i\pi/4} (H S^\dagger)^2 = S H$. Up to an overall phase, we arrive at the rotation gate
\begin{equation}
\label{eq:R}
R = H T H S^\dagger H.
\end{equation}
We were permitted to remove the $S^\dagger$ gate on the right end because this matrix $R$ still rotates the eigenstates of $\sigma_z$ into the eigenstates of $\sigma_\theta$ -- the resulting eigenstates of $\sigma_\theta$ just have different overall phases when the $S^\dagger$ on the right end is removed.  This is clear from the equations $\sigma_\theta = R S^\dagger \sigma_z S R^\dagger = R  \sigma_z  R^\dagger$.

We note that a measurement in the $\theta$ direction when our qubit is in state $\left|\psi\right\rangle$ is given by
\begin{equation}
\begin{aligned}
\left\langle O_\theta\right\rangle &= \left|\prescript{}{\theta}{\left\langle 1 \right|}\left.\psi\right\rangle\right|^2 - \left|\prescript{}{\theta}{\left\langle 0 \right|}\left.\psi\right\rangle\right|^2\\
&= \left|\prescript{}{z}{\left\langle 1 \right|}R^\dagger\left|\psi\right\rangle\right|^2 - \left|\prescript{}{z}{\left\langle 0 \right|} R^\dagger\left|\psi\right\rangle\right|^2.
\end{aligned}
\end{equation}
Thus, to take a $\theta$ measurement, we simply apply $R^\dagger$ to our state, make a $z$ measurement, and then apply $R$ to the result.

Now that we have tailored Fig. \ref{exp} to the IBM 5Q, we can run the Leggett-Garg test. Fig. \ref{parta} gives the circuit for protocol (a). The circuit begins with operation $O_1$ comprised of an $X$ gate that flips state $\left|0\right\rangle_z$ to state $\left|1\right\rangle_z$ and the set of gates $R$ that rotates the state to $\left|1\right\rangle_\theta$.  At the end of protocol (a), the $z$-directional measurement $O_3$ is taken.

Moving on to determine $\epsilon_b$ from protocol (b) and $\epsilon_d$ from protocol (d), we have the two circuits shown in Fig. \ref{partbd}. 

Note the use of the $CNOT$ gate to record the intermediate state on the second qubit. Because the $CNOT$ gates can only have $Q_2$ as the target qubit, we must add $H$ gates directly before and after the application of the $CNOT$ gate to both of the qubit states in each experiment. This causes the target qubit and control qubit to exchange roles.

While Fig. \ref{parta} and \ref{partbd} show exactly which gates are placed in the circuit and exactly where they are placed in the circuit, several additional gates are placed in the IBM 5Q interface to prevent the IBM 5Q complier from changing these circuits during execution.  In protocol (b), for example, we have two Hademard gates in a row, $HH$. To keep the IBM 5Q from collapsing the two gates into an identity gate, we inserted the operator combination $T T^\dagger$ between the two Hademard gates \cite{JayGambetta}.  This $T T^\dagger$ combination prevents the compiler from combining $HH$ into an identity gate but does not have any other effect on the circuit execution since $T$ and $T^\dagger$ gates physically correspond to timing delays rather than actual pulses.  Whenever an instance of $HH$ is found in a protocol, we actually insert $H T T^\dagger H$ into the IBM 5Q interface.

Additionally, the IBM 5Q allows use of the $Id$ gate, or identity gate. To ensure that the IBM 5Q compiler applies the second $H$ gate on $Q_1$ ($Q_4$) directly after the $CNOT$ gate, we  fill the space between the second $H$ gate applied to $Q_1$ ($Q_4$) and the $z$-directional measurement at the end with identity gates. Otherwise, the IBM 5Q compiler would apply the $H$ gate immediately before the final measurements \cite{JayGambetta}.  This technique is used whenever we wish to impose a fixed time interval between a gate operating on a qubit and the final $z$-measurement.

\begin{figure}[h]  
\centering     
		\subfloat[(c)]{
 			\includegraphics[width=14cm]{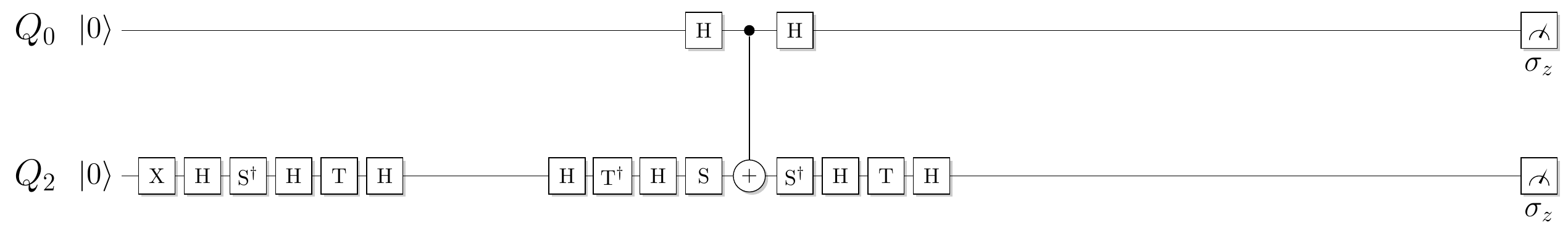} 
		}     
		\vspace{.5cm}
		
		\subfloat[(e)]{
 			\includegraphics[width=14cm]{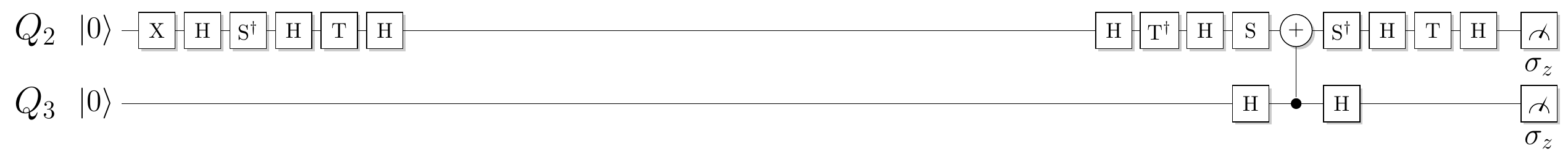} 
		}     
	    \caption{Protocols (c) and (e) of the experiment program implemented in circuits.}
    \label{partce}
\end{figure}

Protocols (c) and (e), shown in Fig. \ref{partce}, contain intermediate measurements in the $\theta$-direction.  By combining these with the results from the circuit in Fig. \ref{parta}, we determine $\epsilon_{c}$ and $\epsilon_{e}$.

In these protocols the instances of $HH$ on either side of the $CNOT$ gates are collapsed into identities. (The same collapse occurs in protocol (f), so that we really are individually testing the operations in (f) as required by our Leggett-Garg program.) This allows for a reduced number of gates necessary in the circuits.

Finally, Fig. \ref{partf} gives us the circuit necessary to measure $\left\langle O_1 O_2\right\rangle_f$ and $\left\langle O_2 O_3\right\rangle_f$. All five qubits are used and $R^\dagger$ and $R$ are applied in an interleaved pattern to alternate measurements back and forth between the $\theta$- and $z$- directions. The qubit $Q_1$ is chosen for the measurement $O_2$ because it has the longest relaxation times of the five qubits of the IBM 5Q and in our experience gave the most reliable results.  (This is one of many specific choices in Fig. \ref{parta} - \ref{partf} and in the definition (\ref{eq:R}) that permitted us to achieve a Leggett-Garg violation.  The fidelity of the gates in the IBM 5Q system is currently too low to achieve a violation for generic implementations of Fig. \ref{exp}.)

\onecolumngrid

\twocolumngrid

\section{Results}
\label{Results}
\begin{figure}[h]   
    \includegraphics[width=14cm]{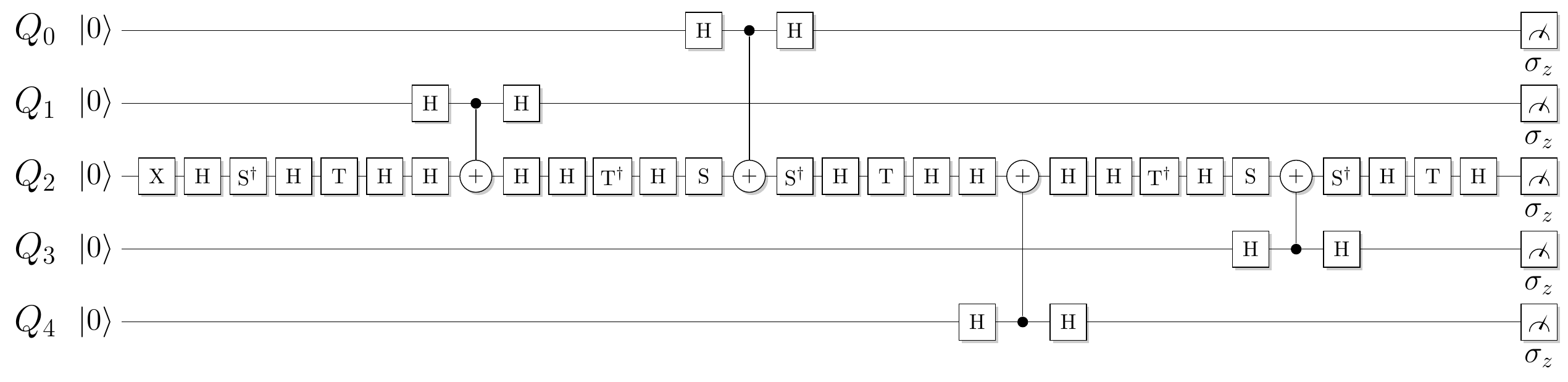}   
    \caption{Protocol (f) of the experiment program implemented in circuits.}
    \label{partf}
\end{figure}

Results are summarized by the tables in Fig. \ref{tab}. We performed 10 repetitions of the complete experimental program, all six protocols given by Figs. \ref{parta}-\ref{partf}.  The data from these 10 repetitions allowed us to compute the error bounds given in the tables. Every time we took a measurement, it was actually the output of $r$ repeated executions of the IBM 5Q hardware, where we set $r=8192$ in the IBM 5Q interface. The Leggett-Garg quantity table gives the average measurements obtained from experiments (a) and (f).  The second table, labelled Adroitness Test Results, gives the correlation function measurements from protocols (b)-(e).   We evaluate the adroitness of each measurement using equations like eq. (\ref{eq:epsilonb}) and total them according to eq. (\ref{eq:epsilontotal}).  For reference, we include the quantum mechanical prediction for each value in the table.

The data confirm that both of the conditions specified in eq. (\ref{eq:bothLG}) are met: the calculated $LG$ is indeed negative and $\left|LG\right| \geq \epsilon_{total} $. 

\onecolumngrid

\twocolumngrid

\onecolumngrid

\begin{center}
\begin{figure}[h]
\centering
\begin{tabular}{|c|c|c|c|c|}
\hline
\multicolumn{5}{|c|}{\textbf{The Leggett-Garg Quantity}} \\
\hline
&$\left\langle O_1 O_3\right\rangle_a$ & $\left\langle O_1 O_2\right\rangle_f$ & $\left\langle O_2 O_3\right\rangle_f$ & $LG$\\
\hline
\hline
Measured & $-0.70 \pm .01$ & $-0.69 \pm .01$ & $0.18 \pm .02$ & $-0.21 \pm .03$\\
\hline
\hline
$\begin{array}{c}\text{Quantum}\\ \text{Prediction}\end{array}$ & $-\frac{1}{\sqrt{2}} \approx -0.70$ & $-\frac{1}{\sqrt{2}} \approx -0.70$ & $\frac{1}{4} = 0.25$ & $-\sqrt{2} + \frac{1}{4} +1 \approx -0.16$ \\
\hline
\end{tabular}
\\
\vspace{0.5in}
\begin{tabular}{|c|c|c|c|c|c|}
\hline
\multicolumn{6}{|c|}{\textbf{Adroitness Test Results}} \\
\hline
& $\left\langle O_1 O_3\right\rangle_{b}$&$\left\langle O_1 O_3\right\rangle_{c}$&$\left\langle O_1 O_3\right\rangle_{d}$&$\left\langle O_1 O_3\right\rangle_{e}$&$\epsilon_{total} $\\
\hline
\hline
Measured & $-.69 \pm .02$ & $-0.71 \pm .02$ & $-0.68 \pm .01$ & $-0.67 \pm .02$ & $.08 \pm .04$\\
\hline
\hline
$\begin{array}{c}\text{Quantum}\\ \text{Prediction}\end{array}$ & $-\frac{1}{\sqrt{2}} \approx -0.70$ & $-\frac{1}{\sqrt{2}}\approx -0.70$ & $-\frac{1}{\sqrt{2}} \approx -0.70$  & $-\frac{1}{\sqrt{2}} \approx -0.70$  & 0 \\
\hline
\end{tabular}
\caption{The Leggett-Garg result with adroitness test results.  Experimentally measured results show agreement with the predictions of quantum mechanics.}
\label{tab}
\end{figure}
\end{center}

\twocolumngrid

\section{Conclusion}
\label{Conclusion}

We have carefully framed a Leggett-Garg program that (a) addresses the clumsiness loophole and (b) is suited for execution on the IBM 5Q Quantum Experience.  This program demonstrated that qubit $Q_2$ of IBM 5Q is not a macrorealistic system being measured noninvasively.  It also supplies compelling evidence that noninvasiveness in the measurements does not exclusively derive from mundane experimental clumsiness.  This suggests that it is impossible to formulate a noninvasive macrorealistic description of $Q_2$.

Some recent papers have stressed the role of equalities rather than inequalities in testing macrorealism \cite{Kofler2013,Clemente2016}.  One might consider reframing our Leggett-Garg program in the future by directly checking the equality (\ref{eq:correlatorequality}) rather than inserting it into the inequality (\ref{eq:LG}).

\section*{Acknowledgments}
We are grateful to Ben Palmer for suggesting that we analyze the quantum character of the IBM 5Q.  We acknowledge use of the IBM Quantum Experience for this work. The views expressed are those of the authors and do not reflect the official policy or position of IBM or the IBM Quantum Experience team.

\bibliographystyle{apsrev4-1}
\bibliography{ref}

\end{document}